\documentclass{article}

\usepackage{PRIMEarxiv}

\usepackage[utf8]{inputenc} 
\usepackage[T1]{fontenc}    
\usepackage{url}            
\usepackage{booktabs}       
\usepackage{amsfonts}       
\usepackage{amsmath}        
\usepackage{nicefrac}       
\usepackage{microtype}      
\usepackage{xcolor}         
\usepackage{lipsum}
\usepackage{natbib}
\usepackage{fancyhdr}       
\usepackage{graphicx}       
\graphicspath{{media/}}     
\usepackage{algorithm}
\usepackage{algpseudocode} 
\usepackage{verbatim} 
\usepackage{amsfonts} 
\usepackage{hyperref} 
\hypersetup{
    colorlinks=true,
    urlcolor=blue
}

\title{ORCAS - Orthogonal Representation for Compression of distributed Acoustic Sensing
}

\author{
  Roman Pavelkin, Luis A. Zavala-Mondragon, Fons van der Sommen \\
  Eindhoven University of Technology \\
  Eindhoven\\
  \texttt{\{r.pavelkin, l.a.zavala.mondragon, fvdsommen\}@tue.nl} \\
}

\begin{document}
\maketitle

\begin{abstract}

Distributed Acoustic Sensing (DAS) turns standard fiber-optic cables into dense virtual sensor arrays. Using Rayleigh backscattering of coherent laser pulses, DAS measures vibrations and strain over 100 km with meter-scale resolution and is well suited to remote seismic sensing, pipeline leak detection, and infrastructure protection, where conventional point sensors are too costly or impractical. This high temporal and spatial resolution generates massive data volumes, creating major processing and storage challenges that demand optimized compression algorithms. This paper introduces \textbf{O}rthogonal \textbf{R}epresentation for \textbf{C}ompression of distributed \textbf{A}coustic \textbf{S}ensing~(\textbf{ORCAS}), a framework for rapid and efficient compression of DAS signals. The method transforms dense DAS data into a sparse representation via a learned orthogonal transform. In contrast to iterative sparse coding (e.g. orthogonal matching pursuit or least-angle regression), ORCAS provides swift sparse encoding with throughput exceeding 250 MB/s via simple projection and top-K thresholding. Another key element of the ORCAS framework is the DAS-tailored structural-entropy coding scheme for the sparse signal, which employs the Lloyd-Max algorithm to optimize the quantization thresholds according to the underlying data distribution, thereby improving data fidelity even under strong compression. As a result, the proposed system supports real-time total throughput exceeding 90 MB/s and attains a superior rate–distortion performance relative to the established industrial compression standards, such as JPEG2000, ZFP, Zstd, and recently published DASPack, which was demonstrated on the test set built from diverse DAS data patches. With ORCAS, for the higher-compression settings, the achieved compression ratio is approximately 13 with an SNR and PSNR after decompression of 13~dB and 28~dB, respectively. Furthermore, for its higher-end settings, ORCAS reaches compression ratios of approximately 6.5 with SNR and PSNR of 27~dB and 42~dB. Following the principles of reproducibility and open science, all evaluations were conducted on a heterogeneous dataset compiled from DAS measurements obtained from publicly available repositories. The ORCAS implementation is available on our \href{https://github.com/RomanPavelkin/orcas}{GitHub repository}.

\end{abstract}


\section{Introduction}
\label{intro}

Distributed Acoustic Sensing (DAS) is an innovative fiber-optic technology that transforms standard optical fibers into dense, continuous arrays of sensors capable of measuring vibrations and strain~\cite{parker_distributed_2014, zhan_distributed_2020}. Using short light pulses and analyzing Rayleigh backscattering, DAS achieves high spatial resolution—down to the sub-meter scale—and rapid sampling rates over long distances~\cite{parker_distributed_2014, he_optical_2021, gorshkov_scientific_2022}. The referred setting enables diverse applications~\cite{he_optical_2021} in engineering, geophysics, material science, biology, and humanitarian efforts, most notably in seismology to monitor ground motion and utilize existing fiber optic infrastructure, thus reducing costs~\cite{lindsey_fiberoptic_2017, dou_distributed_2017, ajo-franklin_distributed_2019, lindsey_fiber-optic_2021}.

In spite of its advantages, DAS generates massive data volumes —up to one terabyte per fiber daily— posing challenges in data transmission, storage, and processing~\cite{markom_systematic_2025, rafi_big_2026, van_der_horst_-well_2013}. Although storage and transmission challenges could be partly mitigated with efficient signal compression schemes, it should be noted that DAS data is heterogeneous. Moreover, environmental factors can further complicate compression efforts~\cite{ni_wavefield_2024, dong_real-time_2022}. In order to mitigate the challenges related to DAS storage and transmission, we could refer to conventional compression methods, such as GZIP and ZFP. However, these methods provide limited efficiency~\cite{song_using_2022, shang_research_2022, dong_real-time_2022}, because their original design does not explicitly account for the characteristic spatio-temporal structure of fiber-optic strain measurements.  However, some improvements have been partly addressed by newer approaches, such as H5TurboPFor and DASPack~\cite{segui_daspack_2026}, which have improved storage reduction while keeping the signal distortion at an acceptable level.

Recent advancements in signal compression consider the inclusion of machine learning within the pipelines. For example, Recurrent AutoEncoders~(RAE)~\cite{chiarot_rat-cc_2025} and SHallow REcurrent Decoders (SHRED)~\cite{williams_sensing_2024} perform well in wavefield reconstruction and low-frequency ocean waves compression, though with limited generalizability and practical use, as they respond poorly to varying conditions~\cite{ni_wavefield_2024, wang_deep_2024, chen_deep_2025}. This problem is further complicated the lack of standardized benchmarks for compression quality within the DAS community, which challenges ranking algorithms under common conditions.

To tackle the challenge of compressing the general DAS measurements, the Orthogonal Representation for Compression of distributed Acoustic Sensing (ORCAS) system has been developed, yielding a 13× compression ratio with acceptable levels of distortion (SNR of 13~dB and PSNR of 28~dB). ORCAS employs a fast orthogonal transformation tailored to DAS data, followed by nonlinear quantization and structural-entropy coding. A comprehensive analysis of the ORCAS algorithm is provided, revealing the system's promising utility for both machine learning practitioners and DAS specialists in various applications, including seismological research and infrastructure monitoring.

\section{Related work on DAS compression}
\label{sect:rel_work}

The increasing data volumes from DAS systems have led to significant research in compression techniques. Established methods, like JPEG, which is based on the Discrete Cosine Transform (DCT), and JEPG2000, based on wavelet decomposition, leverage spatio-temporal coherence for efficient data representation and are fundamental to codecs used for seismic data~\cite{ni_wavefield_2024, sebai_seismic_2024}. Alternatively, predictive coding techniques, including Differential Pulse-Code Modulation (DPCM), help to reduce bitrates with minimal computational needs~\cite{wood_seismic_1974} by exploiting the high inter-channel and inter-sample correlation typical in DAS measurements. In contrast, general-purpose lossless compression algorithms like GZIP and Zstd, which operate on repetition and redundancy, are prevalent in the DAS field~\cite{deutsch_deflate_1996, collet_zstandard_2018, dong_real-time_2022}. Furthermore, more specialized methods, such as H5TurboPFor and ZFP, are suited for lossless and lossy compression of different data types~\cite{lindstrom_fixed-rate_2014}. More recently, specialized DAS compression tools, such as DASPack, offer a versatile open-source framework with a three-stage compression pipeline, achieving significant compression ratios while facing performance challenges at higher compression levels~\cite{segui_daspack_2026}.

Recent advancements in machine learning, particularly autoencoder architectures, show promise for better compression by learning data representations directly. The Recursive AutoEncoder~(RAE) exhibits better performance than traditional methods based on DCT and wavelet transform~\cite{wang_deep_2024}, while a Visual Transformer (ViT)-based approach also shows competitive results when compared, for example, with the wavelet-transform-based method and CNN~\cite{chen_deep_2025}. Note that both of the referred deep learning solutions face generalizability issues and are resource-intensive~\cite{wang_deep_2024, chen_deep_2025}, limiting real-time application.

\section{Materials and Methods}
\label{sect:mat_met}

\subsection{Dataset composition}
\label{sect:mat_met_1}

One of the key challenges in developing machine learning algorithms for distributed acoustic sensing~(DAS) applications is the heterogeneity of available data--such as measurements made using different interrogators, channel spacing, sampling frequency, etc--a limitation recognized in existing literature~\cite{zhu_seismic_2023, lapins_-n2n_2024, dumont_deep_2020, tejedor_machine_2017}. In order to design a robust compression algorithm for various DAS applications, we accessed PubDAS~\cite{spica_pubdas_2023}~-~a public DAS datasets repository for geosciences hosting large volumes of DAS measurements. Specifically, we took records from the PubDAS repository collected during the Global DAS Month~\cite{wuestefeld_global_2024}, when DAS measurements were collected simultaneously in different locations across the globe. For our study, the local measurements were selected from Indonesia (AP\_Sensing-Traintrack, DS\_DAT20230212\_EQ\_Indonesia), Romania (AP\_Sensing-Traintrack, DS\_DAT20230214\_EQ\_Romania), Istanbul (ETH\_Zurich-Istanbul), Iceland (GFZ-NorthIceland), Austria (TUGraz-GrazCity), Svalbard (NTNU\_Svalbard), and Alaska (SNL\_Alaska), ensuring global representation across various geophysical conditions (marine, urban, seismic active areas) and various measurement set-ups (spatio-temporal resolution, measurement units (strain or strain rate)). Detailed descriptions of these measurements are available in the corresponding publication~\cite{wuestefeld_global_2024}; Table~1 summarizes the key characteristics of each DAS measurement included in the study. 

Training and validation sets were organized using data from AP Sensing, ETH (Istanbul), GFZ (Iceland), and TU Graz; the sets included 15,000 and 1,500 preprocessed 2D patches, respectively. The details of the DAS data preprocessing used in this study is given in Section~\ref{sect:mat_met_2}. The test set, consisting of 6,000 patches, was created from NTNU CGF and SNL (Alaska) datasets to prevent data leakage and assure the reliability of performance evaluations. The datasets employed a hybrid sampling strategy of 70\% informative patches (high mean energy) and 30\% random patches, focusing on event emphasis while covering background variability for machine learning applications. These patch selections will be made publicly available upon the study's publication. 

\begin{table}[h]
\caption{Summary of the DAS measurements selected for this study. Columns indicate, respectively, the data providing organization; interrogator manufacturer and model; Channel spacing (Ch. Sp.); the total cable length (Cab. L.); number of channels (\textnumero~Ch.); gauge length (G. L.); and temporal sampling rate (Samp. Rate).} 
\label{tab:fonts}
\begin{center}       
\begin{tabular}{|l|l|l|l|l|l|l|}
\hline
\rule[-1ex]{0pt}{3.5ex}  Provider (Site) & Instrument & Ch. Sp. (m) & Cab. L. (m) & \textnumero~Ch. & G. L. (m) & Samp. Rate (Hz)  \\
\hline\hline

\rule[-1ex]{0pt}{3.5ex}  AP Sensing & AP Sensing & 19.6 & 44500 & 2270 & 10 & 100  \\
\hline

\rule[-1ex]{0pt}{3.5ex}  ETH (Istanbul) & Silixa iDAS & 16 & 8000 & 500 & 10 & 100  \\
\hline

\rule[-1ex]{0pt}{3.5ex}  GFZ (Iceland) & Silixa iDAS & 1 & 5500 & 400 & 10 & 1000  \\
\hline

\rule[-1ex]{0pt}{3.5ex}  TU Graz & FEBUS A1-R & 9.6 & 2600 & 271 & 20 & 100  \\
\hline\hline

\rule[-1ex]{0pt}{3.5ex}  NTNU CGF & ASN OptoDAS & 24.5 & 50000 & 2040 & 8.2 & 625  \\
\hline

\rule[-1ex]{0pt}{3.5ex}  SNL (Alaska) & Silixa iDAS & 24.6 & 37096 & 1509 & 10 & 1000  \\
\hline

\end{tabular}
\end{center}
\end{table}

\subsection{DAS standardization preprocessing pipeline}
\label{sect:mat_met_2}

\begin{figure}
    \begin{center}
    \begin{tabular}{c}
    \includegraphics[width=1.0\linewidth]{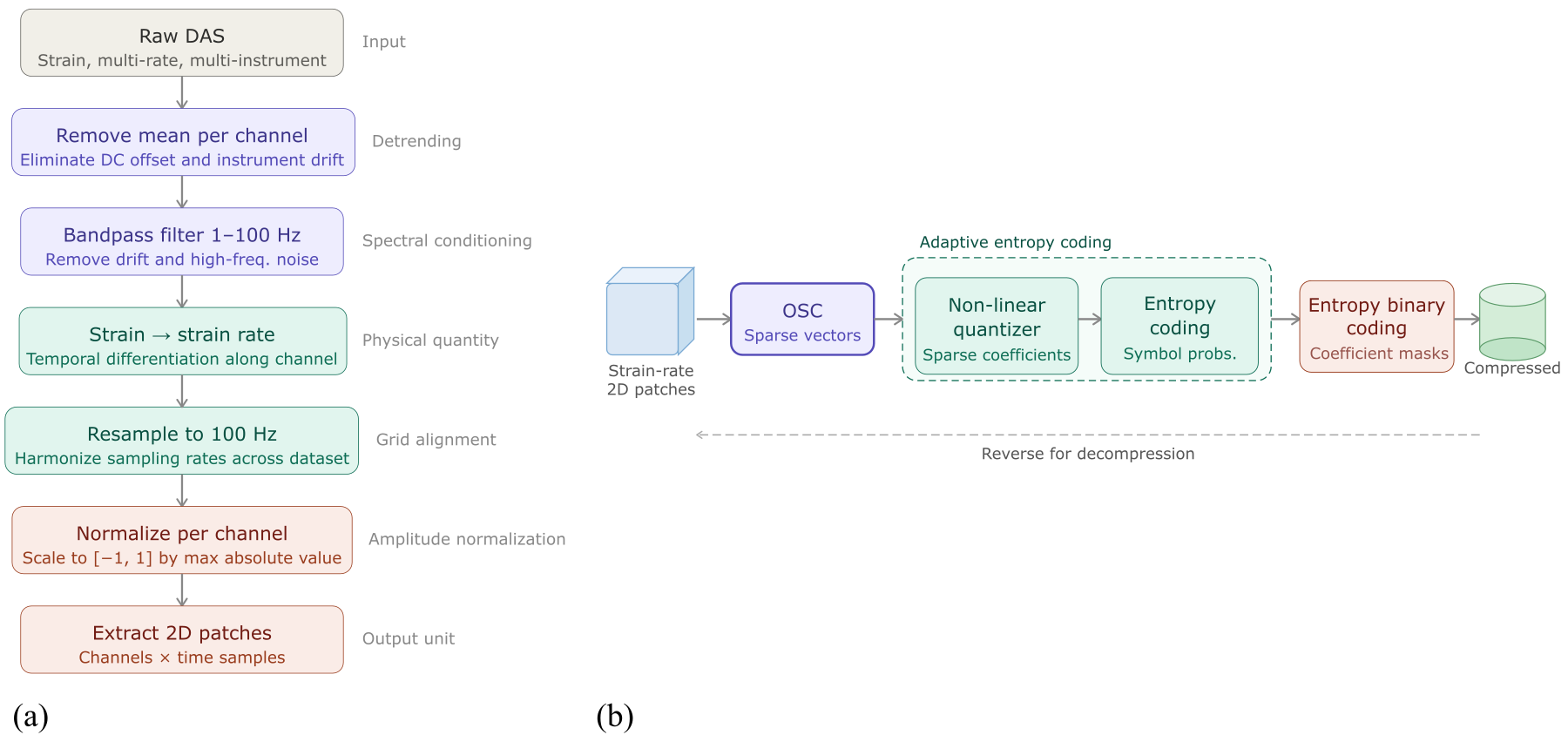}
    \end{tabular}
    \end{center}
    \caption{(a) DAS data preprocessing workflow; (b) ORCAS compression scheme: OSC - Orthogonal Sparse Coding; bin. struct. - binary structural.} 
    \label{fig:preproc_orcacs_workflows}
\end{figure}

To standardize the diverse dataset, all raw DAS measurements undergo a seven-stage preprocessing pipeline before extracting patches (see Fig.~\ref{fig:preproc_orcacs_workflows}~(a)). Initially, DC offset is removed by subtracting each channel's temporal mean, eliminating biases from instrument drift. Afterwards, bandpass filtering is applied to remove low-frequency drift and high-frequency noise~\cite{zhan_distributed_2020, gorshkov_scientific_2022}. The filtering range in our experiments was from 1 Hz to 100 Hz--that spectral content is of interest for DAS applications in seismology, oceanography, and infrastructural monitoring~\cite{gorshkov_scientific_2022, rafi_big_2026}. DAS measurements are acquired in strain or strain rate units, with the latter favored for machine learning due to its suppression of drift and enhancement of transient phenomena~\cite{becker_distributed_2019}. The final steps include per-channel resampling of the signal to a common frequency (100~Hz) followed by amplitude normalization to a range of [-1, 1]. The normalized 2D (time, length) DAS measurement is then split into 2D patches of size 64$\times$16 for efficient processing and localized context for the encoder. The aforementioned patching strategy was chosen because DAS data is inherently spatiotemporal: many physical phenomena, such as seismic waves, vehicles, and whale calls, appear as coherent patterns across channels, creating slanted or structured patterns in the space-time image. Consequently, an ML model operating on 2D patches is capable of identifying and leveraging these underlying patterns. Because many DAS signals exhibit local coherence—such that wavefronts remain discernible even within relatively small spatial windows—the selection of patch size implicitly defines a trade-off between the fidelity with which spatial structure is represented, the resulting dictionary dimensionality, and the computational efficiency of sparse inference. The proposed DAS standardization and preprocessing pipeline can be adapted to the requirements of specific DAS applications. For instance, certain use cases—such as DAS for bioacoustic monitoring—prioritize the kilohertz-range spectral components of the signal and may therefore necessitate tailored modifications of the processing steps.

\subsection{ORCAS pipeline}
\label{sect:mat_met_3}

The operation scheme of the proposed compression method is shown in Fig.~\ref{fig:preproc_orcacs_workflows}~(b). It includes two main stages: (1)~sparsifying transform of the input data (DAS 64$\times$16 patches) and (2)~subsequent adaptive structural-entropy coding of the sparse vectors.

The core data encoder in our method is a learned transform of the input dense DAS signals into sparse vectors--Orthogonal Sparse Coding (OSC), which was introduced in the paper "\textit{Learning efficient data representations with orthogonal sparse coding}"~\cite{schutze_learning_2016}. OSC is based on stochastic descent by Hebbian-like updates and Gram–Schmidt orthogonalizations. Dictionaries obtained through OSC can reconstruct the underlying generating basis even for relatively low and, strikingly, a priori unknown sparsity levels, while remaining tailored to the given signal class~\cite{schutze_learning_2016}.

For our application, we modified the original orthogonal dictionary learning algorithm by incorporating an accelerated training procedure. To this end, we implement mini-batch training in contrast to the original algorithm, which leads to a significant reduction in training time. In addition, we do not set a specific sparsity level during training. We chose this training strategy because in the original paper, the authors demonstrated that the difference in final performance between a model trained with a constrained sparsity level and a model trained without such a constraint~(full~OSC) is insignificant (see, for example, Fig.~7 in the original work by Sch\"utze~et~al.~\cite{schutze_learning_2016}). Algorithm~\ref{alg:orth_dict_learn} outlines our implementation of the orthogonal dictionary learning. In our learning setting, $\varepsilon_{\mathrm{init}}=10^{-1}$, $\varepsilon_{\mathrm{final}}=10^{-3}$, $B=256$, $T=100$~epochs.

\begin{algorithm}[t]
\caption{Mini-batch Orthogonal Dictionary Learning}
\label{alg:orth_dict_learn}
\begin{algorithmic}[1]
\Require Data $X \in \mathbb{R}^{N \times L}$, iterations $T$, batch size $B$,
         learning rates $\varepsilon_{\mathrm{init}},\varepsilon_{\mathrm{final}}$
\Ensure  Orthogonal dictionary $U \in \mathbb{R}^{L \times L}$
\State Initialise $U$ orthogonal via QR of a random Gaussian matrix
\For{$t = 1,\ldots,T$}
    \State $\varepsilon_t \leftarrow \varepsilon_{\mathrm{init}}(\varepsilon_{\mathrm{final}}/\varepsilon_{\mathrm{init}})^{t/T}$
    \State Sample batch $X_b \in \mathbb{R}^{B\times L}$ from $X$; \; $X_{\text{res}} \leftarrow X_b$
    \State Rank atoms $\sigma$ by mean squared projection $\|X_b U_{:,j}\|^2/B$, descending
    \For{$i = 1,\ldots,L$}
        \State $h \leftarrow \sigma_i$;\; $u \leftarrow U_{:,h}$;\; $p \leftarrow X_b u$
        \State $u \leftarrow u + \varepsilon_t\big(X_b^{\top}p - \|p\|^2 u\big)$
               \Comment{Hebbian update}
        \State $u \leftarrow u - U_{:,\sigma_{1:i-1}}U_{:,\sigma_{1:i-1}}^{\top}u$ \;(if $i>1$)
               \Comment{orthogonalise}
        \State $u \leftarrow u + \varepsilon_t X_{\text{res}}^{\top}(X_{\text{res}}u)$;\quad
               $u \leftarrow u/\|u\|$
               \Comment{residual update \& normalise}
        \State $X_{\text{res}} \leftarrow X_{\text{res}} - (X_{\text{res}}u)u^{\top}$;\quad
               $U_{:,h} \leftarrow u$
               \Comment{deflate \& store}
    \EndFor
\EndFor
\State \Return $U$
\end{algorithmic}
\end{algorithm}

The coefficients of sparse representation vectors provided by the method described in the previous paragraph follow a Laplacian distribution~\cite{mallat_theory_1989, simoncelli_statistical_1997} and can be statistically optimally quantized using the Lloyd-Max algorithm~\cite{max_quantizing_1960, lloyd_least_1982}. This algorithm minimizes mean squared quantization error by allocating quantization intervals non-uniformly, enhancing quantization fidelity. Following quantization, Huffman entropy coding~\cite{huffman_method_2007} compresses the symbol sequences, exploiting the non-uniform probability distribution produced by the Lloyd-Max quantizer. This combined approach maximizes rate-distortion efficiency. Additionally, compression of non-zero coefficient positions is achieved using the \textit{bitarray} library~\cite{schnell_bitarray_nodate}, which focuses on representing sparse positions efficiently, yielding significant savings in storage compared to naive methods. Overall, this compression strategy effectively handles the structural sparsity present in the data.

\section{Results and Discussion}
\label{sect:res}

\begin{figure}
    \begin{center}
    \begin{tabular}{c}
    \includegraphics[width=0.7\linewidth]{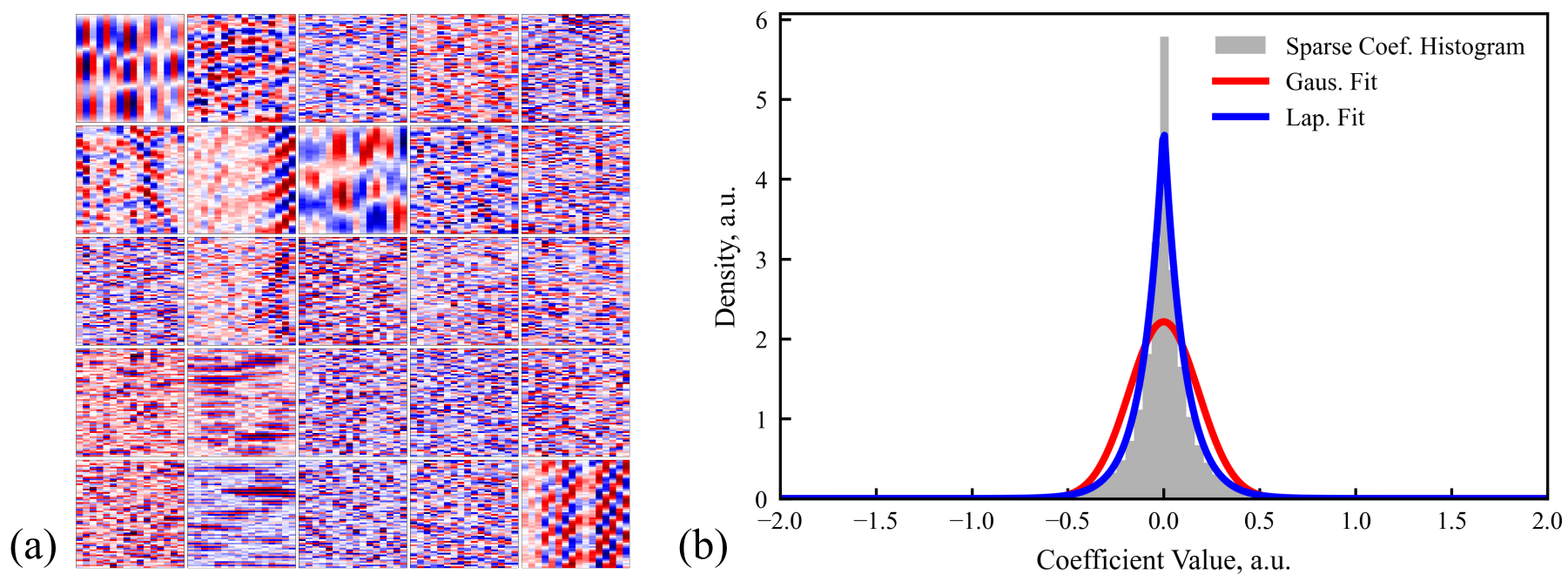}
    \end{tabular}
    \end{center}
    \caption{(a) Gallery of the learned orthogonal DAS atoms (selection of 25 random atoms out of 1024, aspect ratio is set to 1:1); (b) the best Gaussian and Laplacian fits to the distribution of the coefficients of the learned transform.} 
    \label{fig:gal_fits}
\end{figure}

We first present the atoms of the learned orthogonal dictionary, followed by a statistical characterization of the distribution of the coefficients obtained from the transformed test DAS patches (see Fig.~\ref{fig:gal_fits}). 
From Fig.~\ref{fig:gal_fits}~(a), it is apparent that a subset of the learned atoms distinctly represents omnidirectional 2D wave phenomena present in the DAS data, whereas other atoms predominantly capture high‑frequency components and fine‑scale signal texture. Fig.~\ref{fig:gal_fits}~(b) shows that the distribution of the coefficients of the sparse representation is best described by a Laplace distribution rather than a Gaussian distribution. This observation is instrumental in substantiating the selection of the coding scheme employed for the sparse representation of the DAS data.

\begin{figure}
    \begin{center}
    \begin{tabular}{c}
    \includegraphics[width=0.7\linewidth]{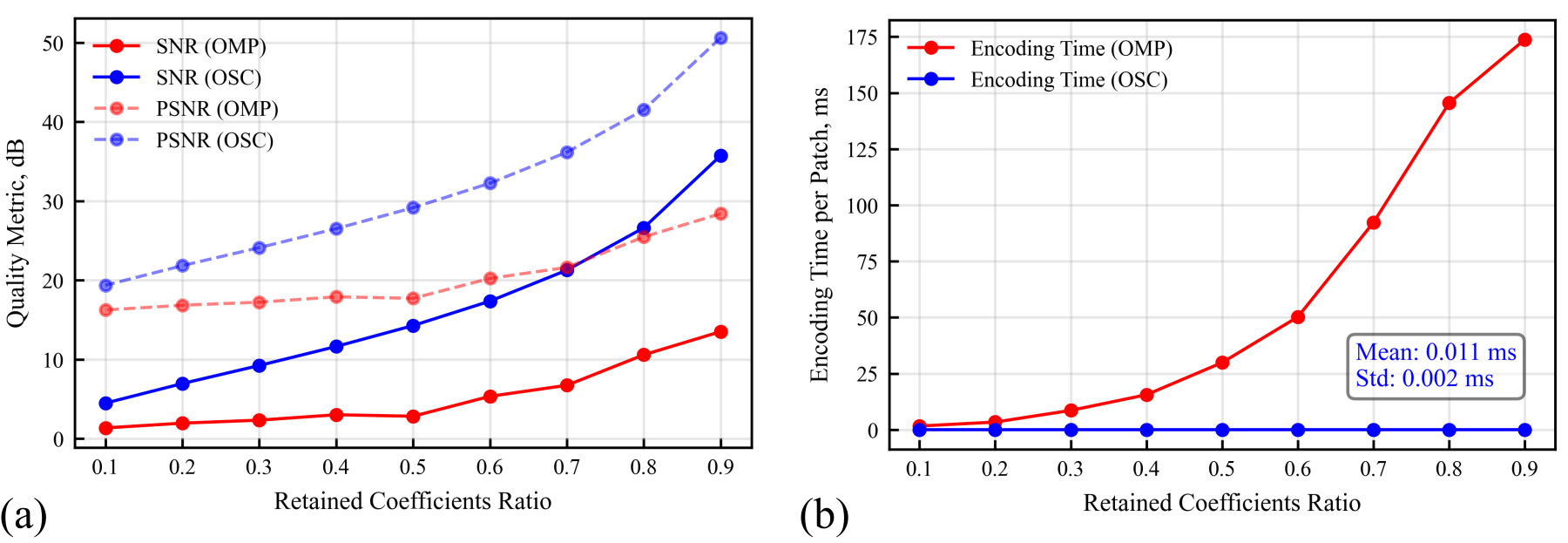}
    \end{tabular}
    \end{center}
    \caption{(a) Average values of SNR and PSNR achieved after sparse encoding/reconstructing the test DAS set using the transforms learned via Orthogonal Matching Pursuit (OMP) and Orthogonal Sparse Coding (OSC) algorithms; (b) average sparse encoding time per data patch $16~\times~64$ pixels.} 
    \label{fig:omp_vs_osc}
\end{figure}

Next, we analyze how the OSC-based encoder performs at different sparsity levels, which are expressed as the ratio of retained coefficients, in comparison with sparse coding via the Orthogonal Matching Pursuit~(OMP) algorithm~\cite{mallat_matching_1993}. To this end, we employed OMP dictionaries, trained separately at each sparsity level. Fig.~\ref{fig:omp_vs_osc}~(a) shows the curves of the reconstruction quality metric~(SNR and PSNR) for patches after encoding and reconstruction.
OSC outperforms OMP significantly in signal reconstruction accuracy, achieving better SNR (PSNR) values: 14.28 (29.16)~dB versus 2.83 (17.72)~dB at a ratio of 0.5, and 4.49 (19.37)~dB versus 1.36 (16.25)~dB at a ratio of 0.1. OSC's orthogonality property contributes to its faster transformation speed, averaging 0.011~ms per patch, compared to OMP's 30~ms and 1.6~ms at the respective sparsity levels. Therefore, the DAS signal sparse encoding strategy is both computationally efficient and high-fidelity.

\begin{figure}
    \begin{center}
    \begin{tabular}{c}
    \includegraphics[width=0.38\linewidth]{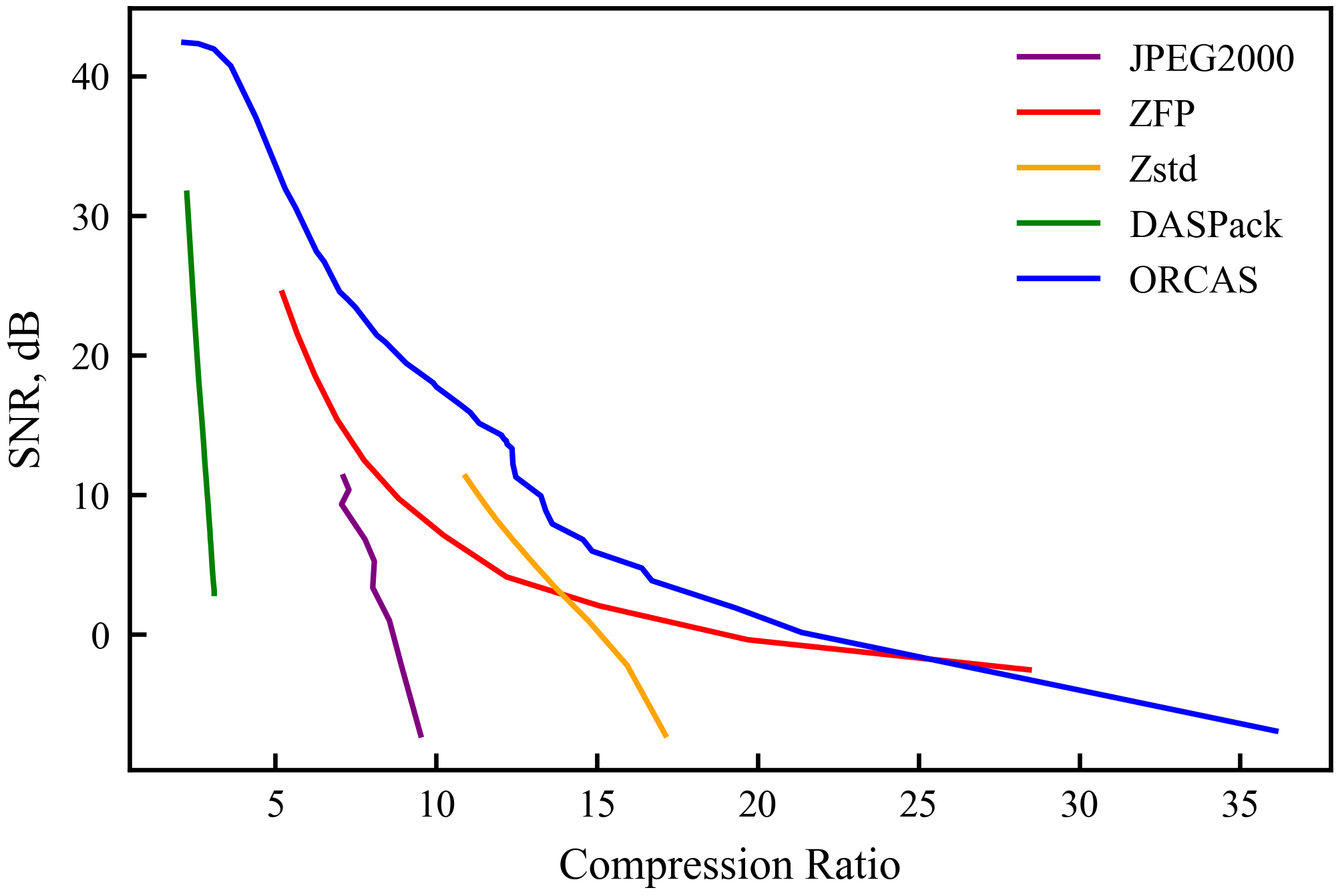}
    \end{tabular}
    \end{center}
    \caption{Rate-distortion curves of the investigated baseline compression algorithms compared to the rate-distortion curve of ORCAS made on the test set of selected DAS patches.} 
    \label{fig:rd_study}
\end{figure}

Finally, we study the rate–distortion characteristics of ORCAS and compare them with those of the baseline compression algorithms (see Fig.~\ref{fig:rd_study}). The degree of compression of ORCAS is tuned by setting 2 parameters: the number of quantization levels and the sparsity level in OSC.
We can see that the curve for ORCAS lies to the right and above the curves for the other methods, which demonstrates that ORCAS provides better quality after decompression at higher compression ratios. Of all the methods, only ZFP approaches the efficiency of ORCAS at the compression ratio around $\times$25. In the balanced regime at the sparsity level of 0.62 with 16 quantization levels, ORCAS achieves SNR~=~13.32~dB (PSNR~=~28.34~dB) and a compression ratio of 13 on the test set. When compressing entire files, ORCAS achieved comparable performance: average SNR values were 14.04~dB and 12.03~dB with compression ratios of 12.06 and 13.67 for the DAS recordings from Svalbard and Alaska, respectively. 

Noteworthy is the throughput of the compression pipeline, since this is a key characteristic for real-time settings. We identified the entropy coding stage as the principal bottleneck in ORCAS throughput and observed that its execution time scales approximately inversely with the compression ratio, following an empirical relationship of the form $2.83/x$, where $x$ denotes the compression ratio. This behavior can be explained as follows: higher compression ratios correspond to a reduced number of non-zero coefficients (i.e., sparser coefficient vectors) and/or a reduced number of quantization levels. Both effects shorten the symbol sequences processed per data block, thereby accelerating Huffman coding and improving overall throughput. For ORCAS in the balanced regime, the total throughput, measured on our machine (Intel(R) Core(TM) i9-13900H, 2.6 GHz, 32 GB of RAM, NVIDIA GeForce RTX 4070, 8 GB of GPU memory), was 93~MB/sec; the OSC transform throughput was 253~MB/sec. These numbers effectively signify that the sparsifying transform method is already fast enough for edge real-time DAS applications; and the current realization of the entropy coding scheme is the major limiting factor in achieving faster performance. However, the demonstrated total throughput can already be sufficient for certain real-time DAS applications ($\geq$50~MB/sec).

\section{Conclusion}
\label{sect:conc}

In this paper, we present ORCAS--a method designed for robust near–real-time compression of distributed acoustic sensing data. From the outset, we aimed to make our algorithm data-tailored. To this end, we trained a machine learning model on a diverse dataset of DAS measurements; the model transforms incoming DAS signals into a sparse representation that can be compressed with high efficiency with the designed structural-entropy encoder. Compared to other compression baselines used in this setting, ORCAS stands out due to its strong generalization to new data and its competitive compression speed.

From the limitations of our study, we can highlight the following: 1)~the still existing bottleneck in compression speed at the stage of entropy coding of sparse data; 2)~the lack of attention on the impact of distortions introduced after data decompression on the performance of downstream systems, such as anomaly detection, etc. These issues may become the subject of investigation in future work. Furthermore, the current compression pipeline also works on preprocessed signals, which might be disadvantageous if the proposed preprocessing is not adequate for specific applications.

\bibliographystyle{unsrt}  
\bibliography{references} 

\end{document}